\documentclass[aps,pra,twocolumn]{revtex4-2} 
\usepackage{amsfonts,amsmath,amssymb}
\usepackage{graphicx}
\usepackage{hyperref}
\usepackage{xcolor}
\usepackage{xspace}
\usepackage{braket}
\usepackage{float}

\newcommand{\ru}{\affiliation{Institute for Molecules and Materials, Radboud 
University, 
		Heyendaalseweg 135, 6525 AJ Nijmegen, The Netherlands}}%
\newcommand{\joemail}{\email[Email:~]{J.Onvlee@science.ru.nl}}%
\newcommand{\SoneDJ}{S($^1D_2$) }
\newcommand{\SthreePJ}[1]{S($^3P_#1$) }
\newcommand{\invcm}{\ensuremath{\text{cm}^{-1}}\xspace}%

\begin{document}
\title{Intense and controlled beam of S($^1D_2$) atoms} 
\author{Alexandra Tsoukala}\ru%
\author{Saskia Bruil}\ru%
\author{Niek Janssen}\ru%
\author{Saskia Pieters}\ru%
\author{Jolijn Onvlee}\joemail\ru% 
\date{\today}%
\begin{abstract}\noindent% 
	We report the production of an intense and controlled beam of electronically excited sulfur atoms in the $^1D_2$ state using a multistage Zeeman decelerator. Sulfur atoms, generated via photolysis of CS$_2$, are produced in both the ground $^3P_J$ and excited $^1D_2$ states. We demonstrate that both can be manipulated using the decelerator, and that temporal separation between them can be achieved by operating in deceleration mode. This enables the generation of sulfur atom beams with a well-defined velocity, narrow velocity spreads, and an enhanced quantum-state purity. To assess the suitability of the beam for scattering studies, we performed a proof-of-principle elastic collision experiment with \SoneDJ and argon atoms. The observed velocity-map-imaging signal confirms that the \SoneDJ beam density is sufficient for detailed scattering studies. These results form the foundation for future studies of reactive and quenching processes involving \SoneDJ atoms at tunable and well-defined collision energies. 
\end{abstract}
\maketitle%

\section{Introduction}

Sulfur (S) atoms play an important role in a wide range of chemical environments, from combustion and atmospheric processes to interstellar chemistry. In their electronically excited $^1D_2$ state, S atoms can undergo electronic quenching, such that they relax to the ground state via collisions \cite{Schofield:JPCRD8:723, Lara:PCCP19:28555, Williams:PCCP27:3722, Black:JPC82:789, Stout:CPL151:156}, or participate in reactive collisions that lead to the formation of new chemical products \cite{Schofield:JPCRD8:723, Ohashi:CPL220:7, Kuroko:jpca127:4055, Tendo:CPL779:138841, DiGenova:ACS9:844, Hickson:IRPC40:457}. These competing pathways offer valuable insights into fundamental energy transfer mechanisms and chemical reactivity. In particular, the \SoneDJ + H$_2$ reaction serves as one of the prototypical systems for investigating the dynamics of complex-forming or insertion reactions \cite{Hickson:IRPC40:457, Aoiz:JPCA110:12546, Guo:IRPC31:1, Rackham:JCP119:12895, Gonzalez:IRPC26:29}.

To investigate such processes experimentally, the crossed molecular beam technique has proven to be a powerful tool, providing detailed insights into the underlying reaction dynamics and energy transfer mechanisms \cite{Yang:AdvSerPhysChem14:ModTrendChemReactDyn, Brouard:book:tutorials, Casavecchia:RPP63:355, Kohguchi:ARPC98:421, Liu:JCP125:132307, Yang:PCCP13:8112, Naulin:IRPC33:427, Aoiz:PCCP2015, Pan:CSR46:7517}. In these experiments, velocity map imaging (VMI) is commonly used to detect the angular and speed distributions of the collision products in order to extract differential cross sections \cite{Eppink:RSI68:3477}. Generally, the resolution in these experiments, both in terms of collision energy and image resolution, is limited by the properties of the atomic and molecular beams involved. Broad velocity distributions as well as insufficient quantum-state control can obscure subtle features in the scattering dynamics and complicate comparisons with theoretical predictions.

To overcome these limitations and to study sulfur atom collisions in the highest possible detail, it would be favorable to produce an intense and controlled beam of \SoneDJ atoms with a well-defined velocity, narrow velocity spreads, and, when necessary, a high quantum-state purity. Such a beam would allow for precise control over the collision energy and could significantly improve the resolution of VMI images, thereby enhancing the level of detail accessible in scattering experiments. 

In previous crossed-beam scattering studies, beams of \SoneDJ atoms have been generated by ultraviolet (UV) photolysis of supersonically expanded carbon disulfide (CS$_2$), typically seeded in helium or neon \cite{Lee:CPL290:323, Lee:JPCA102:8637, Lee:APB71:627, Berteloite:PRL105:203201, Lara:PCCP13:8127, Lara:JPCA120:5274, Li:JPCA128:10234}. This photolysis process produces S atoms in both the excited $^1D_2$ and ground $^3P_J$ states, with branching ratios that strongly depend on the photolysis wavelength \cite{Kitsopoulos:JCP115:9727,Mank:JCP104:3609,Tzeng:JCP88:1658,Hu:CPC9:422,Warne:JCP154:034302,Li:MP119:1,Waller:JCP87:3261,McGivern:JCP112:5301,Xu:JCP120:3051}. Most scattering studies have employed 193~nm radiation, which is readily available from an ArF excimer laser, \cite{Lee:CPL290:323, Lee:JPCA102:8637, Lee:APB71:627, Li:JPCA128:10234} while others used tunable radiation near 195~nm \cite{Berteloite:PRL105:203201, Lara:PCCP13:8127, Lara:JPCA120:5274} to produce \SoneDJ.   

In this work, we demonstrate the production of a high-intensity, velocity-controlled beam of electronically excited \SoneDJ atoms using a 3-m-long multistage Zeeman decelerator. The sulfur atoms are generated via photolysis of CS$_2$ at 199.65~nm, which efficiently produces both \SoneDJ and ground-state \SthreePJ{J} atoms. We show that both can be manipulated using the decelerator, and that temporal separation between these states can be achieved by decelerating the S atoms. The resulting \SoneDJ beam has a well-defined velocity, narrow velocity spreads, and an enhanced quantum-state purity if needed, making it ideally suited for high-resolution scattering experiments. As a proof of concept, we present an elastic scattering experiment involving collisions between \SoneDJ and argon (Ar) atoms. The observed scattering signals confirm that the beam density is sufficient for detailed collision experiments.

\section{Experiment}
\begin{figure}
	\centering
	\includegraphics[width=\columnwidth]{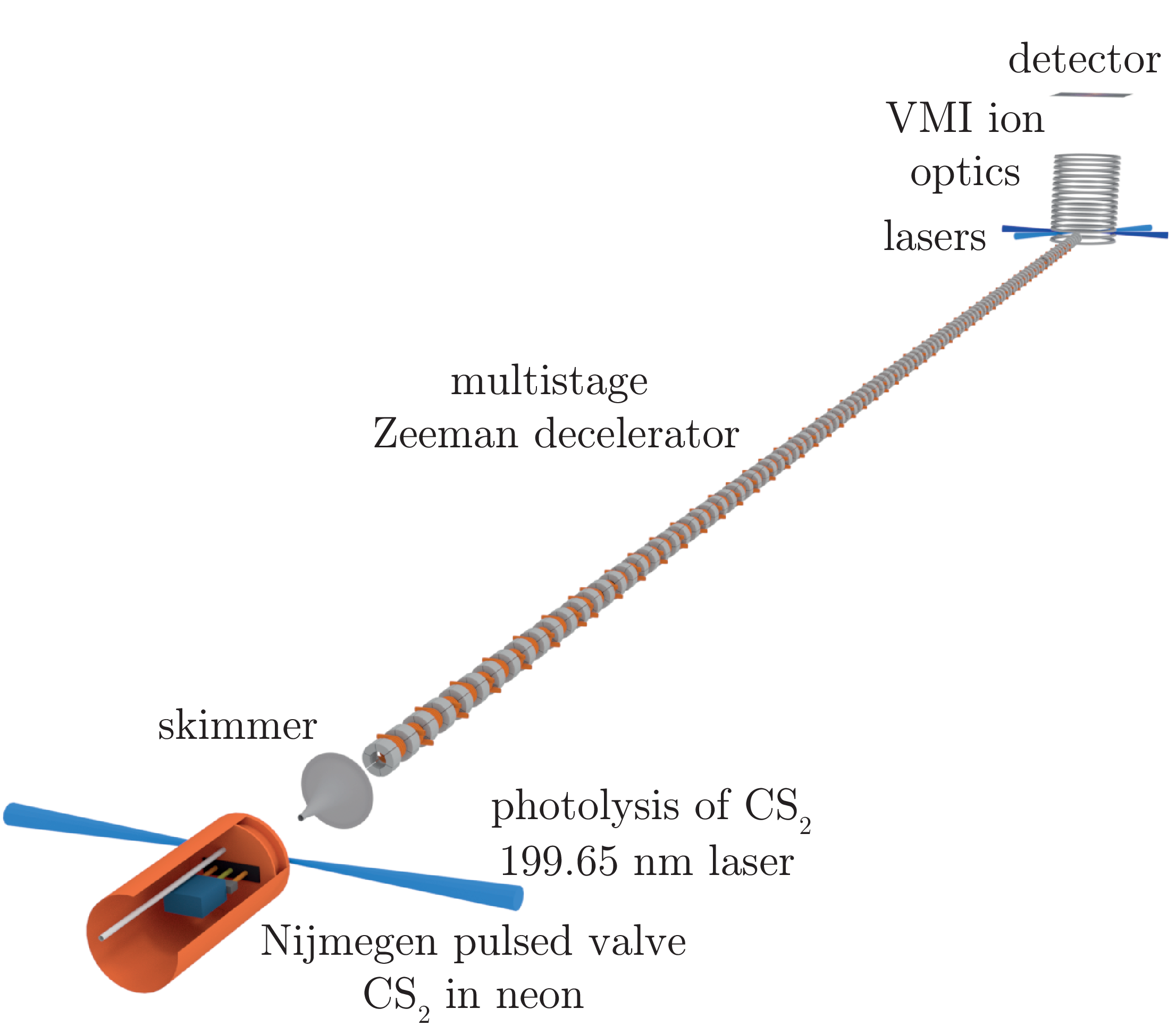}
	\caption{Schematic representation of the experimental setup. CS$_2$ seeded 
	in neon exits a Nijmegen pulsed valve and is photolysed by a 199.65~nm 
	laser. The resulting S atoms pass a skimmer and travel through a multistage 
	Zeeman decelerator that consists of 127 pulsed solenoids and 127 permanent 
	hexapoles. Afterwards, the S atoms are guided into the detection region by 
	7 additional hexapoles and detected using REMPI and VMI.}
	\label{fig:setup}
\end{figure}

The experiments were conducted in a molecular beam apparatus containing a 3-m-long Zeeman decelerator, illustrated in Fig.~\ref{fig:setup}. This machine has been used to 
decelerate beams of O atoms \cite{Cremers:PRA98:033406}, O$_2$ molecules 
\cite{Cremers:PRA98:033406}, NO (X $^2\Pi_{3/2}$) radicals 
\cite{Plomp:JCP152:091103}, NH radicals \cite{Plomp:PRA99:033417}, and C atoms 
\cite{Plomp:JPCL12:12210}. A detailed description of the decelerator can be found elsewhere \cite{Cremers:RSI90:013104}. Here, we focus on the 
experimental aspects related to the production, control, and detection of S atoms. 

\begin{figure}
	\centering
	\includegraphics[width=\columnwidth]{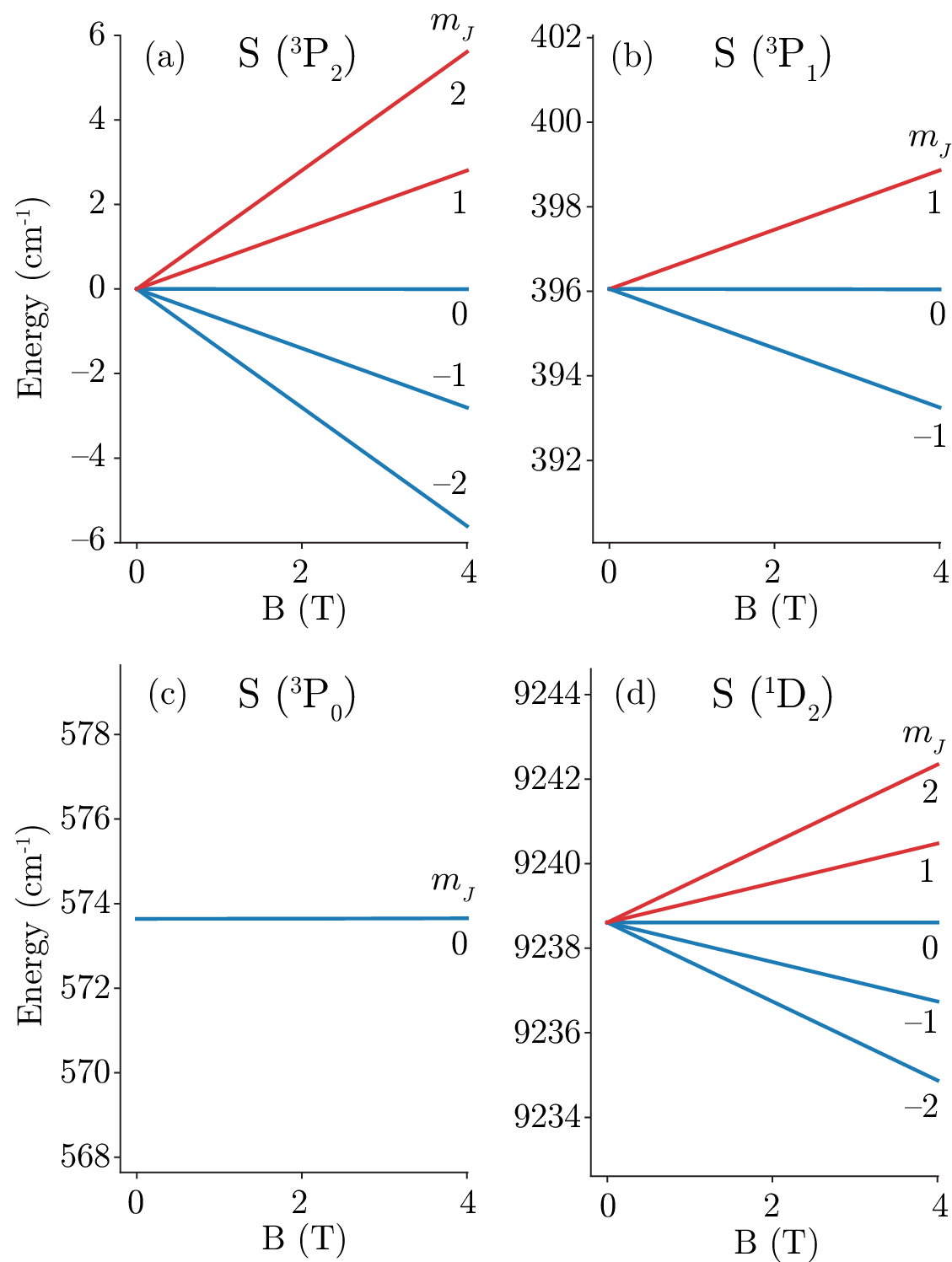}
	\caption{Energy-level diagrams illustrating the Zeeman shifts for S 
	atoms in different states. Low-field-seeking states that can be 
	selected by the decelerator are depicted in red, other states are 
	shown in blue.}
	\label{fig:Zeeman_effect}
\end{figure}

A mixture of 0.5\% CS$_2$ seeded in neon with a backing pressure of 3 bar was expanded through a Nijmegen pulsed valve (NPV) \cite{Yan:RSI84:023102}, operated at a repetition rate of 20 Hz. At the exit of the valve, S atoms were produced by photolysis of CS$_2$ using a pulse energy of 0.45 mJ in a 5~ns pulse of 199.65~nm tunable radiation, focused using a cylindrical lens with 400~mm focal length. This UV radiation was obtained by frequency tripling the output of a dye laser (LIOP-TEC), pumped using the second harmonic of an Nd:YAG laser (InnoLas SpitLight 600). High S-atom signals were found at this wavelength, in agreement with Ref.~\cite{Mank:JCP104:3609}. 
The S atoms with a mean forward velocity of around 875 m/s were formed in both the ground $^3P_{J=2,1,0}$ as well as the excited $^1D_{J=2}$ states. In the $^3P_J$ ground state, the 
sulfur atoms have a magnetic moment of 3, 1.5, and 0 $\mu_B$ for the 
spin-orbit levels $J =$ 2, 1, and 0, respectively. The S atoms in the 
$^1D_2$ excited state have a magnetic moment of 2 $\mu_B$. 
In the presence of a magnetic field, these states split into the different $m_J$ components, where 
$m_J$ is the projection of the total electronic angular momentum $J$ 
on the space-fixed $z$-axis.
The resulting Zeeman energy-level diagrams are shown in Fig.~\ref{fig:Zeeman_effect}.
Sulfur atoms that reside in low-field-seeking $m_J$ components, depicted in red in 
Fig.~\ref{fig:Zeeman_effect}, can be selected by the Zeeman decelerator. 
 Atoms in the $m_J = 0$ components are insensitive to magnetic fields and pass through the decelerator in free flight, which strongly reduces their density and thus their contribution.
 
Approximately 100~mm downstream from the nozzle of the NPV, the S atoms passed a 
3-mm-diameter skimmer and entered the multistage Zeeman decelerator \cite{Cremers:RSI90:013104}, 
consisting of an alternating array of 127 pulsed solenoids 
and 127 permanent hexapoles. Afterwards, the beam was 
guided towards the detection region using 7 additional 
hexapoles. 

To record time-of-flight (TOF) profiles, the S atoms were detected state-selectively using resonance-enhanced multiphoton ionization (REMPI). The \SoneDJ products 
were ionized by (1+1) REMPI via the (4s)$^3S$ state at 216.9~nm {\cite{Weeraratna:CPL657:162}} using a pulse energy of 0.45 mJ in a 5~ns pulse. The \SthreePJ{{2,1,0}} products were 
ionized by (2+1) REMPI via the (4p)$^3P_{2,1,0}$ states using wavelengths between 308 and 311~nm \cite{Hsu:JCP97:6283}, using a pulse energy of 0.14~mJ in a 5~ns pulse. The first laser intersected the molecular beam at $90^\circ$, and the second at $135^\circ$. In both cases, the laser beam was focused in the interaction region by a spherical lens with a focal length of 500~mm. Both laser beams were generated by Nd:YAG (InnoLas SpitLight 1500) pumped dye lasers (LIOP-TEC).
 
The resulting S ions were accelerated by high-resolution VMI ion optics \cite{Plomp:MP119:e1814437} towards a mass-gated microchannel plate detector. The impact positions of the ions were recorded using a phosphor screen in combination with a CMOS camera.

\section{Results and discussion}
\subsection{TOF measurements} \label{sec:TOF}

Within our experimental framework, the decelerator was used in two modes of operation, as described in Ref.~\cite{Cremers:RSI90:013104}. In deceleration mode, each solenoid was pulsed once, leading to a nearly constant reduction of kinetic energy per solenoid and a corresponding decrease in the longitudinal velocity of the S-atom packet. In the so-called hybrid mode, each solenoid was pulsed twice to guide the S-atom packet through the decelerator at a constant speed. In both modes, the final velocity of the packet was determined by the specific pulse sequence used for the solenoids. 

In this study, we focused on final longitudinal velocities around 875~m/s, which can be achieved by seeding in neon. While effective guiding of S atoms is possible in this case, the relatively low magnetic moment-to-mass ratio of S, in combination with the relatively high velocity, limits the amount of deceleration achievable. Other beam velocities could for instance be reached by using a different seed gas. However, using pure helium results in beam velocities that are too high to be efficiently manipulated by the decelerator. On the other hand, heavier rare gases such as argon or krypton produce slower beams, but are known to have significantly higher quenching cross sections for \SoneDJ atoms \cite{Black:JPC82:789}, thereby reducing the \SoneDJ density. Yet, if the beam density is not the main limitation, these heavier seed gasses could be employed to reach lower final velocities.

\begin{figure}
	\centering
	\includegraphics[width=\columnwidth]{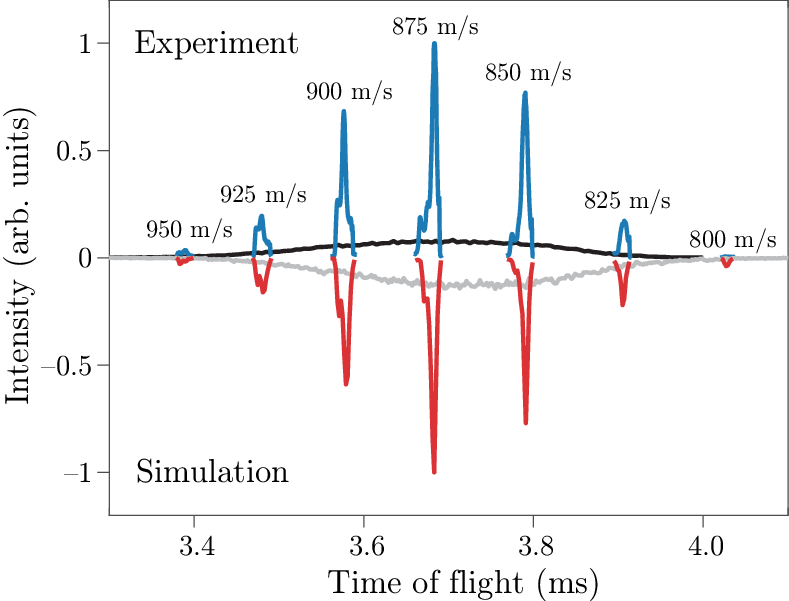}
	\caption{Selected parts of TOF profiles for \SoneDJ atoms exiting the Zeeman decelerator operated in hybrid mode. The measured and simulated profiles are shown in blue and red, respectively. The measured (black) and simulated (gray) profiles obtained when the solenoids are not pulsed are shown for comparison. The experimental profiles are normalized with respect to the profile obtained with the solenoids switched off (black).}
	\label{fig:TOFs}
\end{figure}

Representative TOF profiles of \SoneDJ atoms at various velocities, obtained using the hybrid mode, are shown in blue in the upper part of Fig. \ref{fig:TOFs}. Only the segments of the TOF profiles containing the selected packets are displayed. The final velocity of the packet of \SoneDJ atoms was controlled by selecting a different initial velocity of the beam pulse and guiding this packet at constant speed through the decelerator. For comparison, the TOF profile recorded without pulsing the solenoids, i.e., when the beam is only transversally focussed by the hexapoles, is shown in black. The hybrid mode of the decelerator clearly produces narrow and intense peaks, indicating an effective longitudinal control of the \SoneDJ atoms. 

The measured TOF profiles exhibit good agreement with simulated profiles, shown below the experimental data in Fig.~\ref{fig:TOFs}. These simulated TOFs result from three-dimensional numerical particle trajectory simulations that account for the forces exerted on the S atoms by the space- and time-dependent magnetic fields inside the Zeeman decelerator \cite{Cremers:PRA95:043415, Cremers:RSI90:013104}. The simulations reproduce both the arrival time distributions of the selected packets and the relative intensities of the observed peaks well. This indicates that the properties of the S-atom beam and its transportation through our machine are well understood. 

Since the REMPI scheme used for detecting \SoneDJ atoms imposes around 12~m/s recoil to the S ions, direct extraction of the velocity distributions of the selected \SoneDJ atoms from the experimental data is not feasible. Instead, these can be determined from the numerical particle trajectory simulations. For the profiles shown here, the longitudinal velocity spread is around 7.5~m/s, whereas the transverse spread is around 5~m/s, both given as full width at half maximum (FWHM) values. 

\subsection{Quantum-state distributions}
\begin{figure}[H]
	\centering
	\includegraphics[width=\columnwidth]{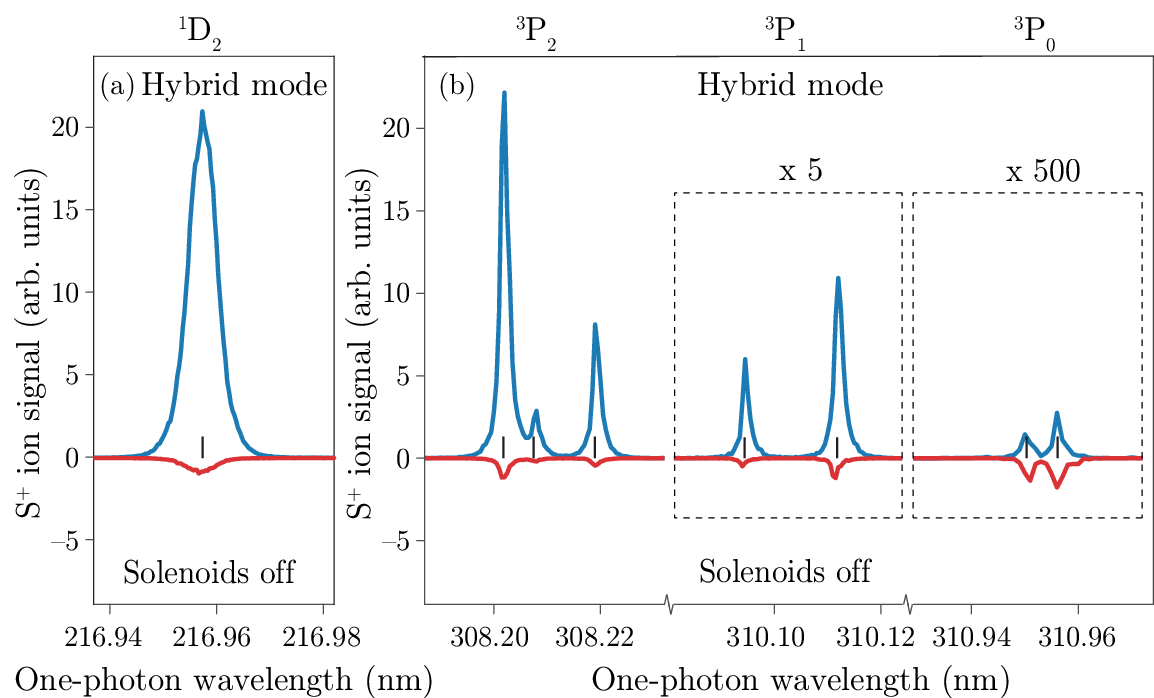}
	\caption{Experimental 1+1 REMPI spectrum for \SoneDJ atoms (a) and 2+1 REMPI spectra for \SthreePJ{{J=2,1,0}} atoms (b) exiting the decelerator. The spectra in the top half (blue) were measured with the decelerator operated in hybrid mode, using a pulse sequence optimized for \SoneDJ atoms. The spectra in the bottom half (red) were measured with the solenoids switched off. The vertical black lines indicate the wavelengths for the resonant transitions to the excited states. All ground-state S-atom signals were measured with the same laser intensity. For clarity, the intensity of the peaks for $^3P_{1,0}$ is magnified 5 and 500 times, respectively.}
	\label{fig:REMPI_1}
\end{figure}

Photolysis of CS$_2$ at 199.65~nm produces S atoms not only in the excited $^1D_2$ state, but also in the ground $^3P_{2,1,0}$ states. This is illustrated in the bottom half of Fig.~\ref{fig:REMPI_1}, which shows measured REMPI spectra for \SoneDJ atoms (a) and ground-state S atoms (b) exiting the decelerator with the solenoids switched off. Note that the intensities for ground- and excited-state S atoms are not directly comparable, since different lasers and REMPI schemes were employed. Each nominal transition probing the \SthreePJ{J} levels contains two or three spectral lines, corresponding to resonant transitions to the different spin-orbit levels of the electronically excited state ($4^3P_{2,1,0}$) \cite{Hsu:JCP97:6283}. These transitions are indicated by vertical black lines in the figure. Around 87\% of the observed ground-state signal originates from the $^3P_2$ state. 

When the decelerator is operated in hybrid mode with a pulse sequence optimized for \SoneDJ atoms, the $^1D_2$ signal increases significantly, see the top half of Fig.~\ref{fig:REMPI_1}a. Since \SthreePJ{J} atoms can also occupy low-field-seeking states, see Fig.~\ref{fig:Zeeman_effect}, they are amenable to magnetic manipulation as well, as shown in the top half of Fig.~\ref{fig:REMPI_1}b. Comparing the top and bottom halves of Fig.~\ref{fig:REMPI_1}b reveals enhanced signals for the $^3P_1$ and $^3P_2$ states, even though the pulse sequence was optimized for \SoneDJ atoms with a magnetic moment of $2~\mu_B$. This increase reflects their interaction with the magnetic fields. In contrast, the $^3P_0$ intensity remains unchanged, reflecting that this state does not interact with magnetic fields, see Fig.~\ref{fig:Zeeman_effect}. Under these conditions, approximately 91\% of the observed ground-state signal can be attributed to the $^3P_2$ state. 

In certain scattering experiments, it is undesirable that the \SoneDJ and \SthreePJ{J} atoms arrive simultaneously in the interaction region. For instance, in studies of \SoneDJ quenching to the ground state, this overlap would lead to contributions of unscattered \SthreePJ{J} atoms that could obscure part of the scattering signal of interest. The use of a Zeeman decelerator enables temporal separation of these states. Here, we focus on separating the \SoneDJ from the \SthreePJ{2} state, which is the most populated ground-state level (see Fig.~\ref{fig:REMPI_1}) and possesses the highest magnetic moment (see Fig.~\ref{fig:Zeeman_effect}).

\begin{figure}
	\centering
	\includegraphics[width=\columnwidth]{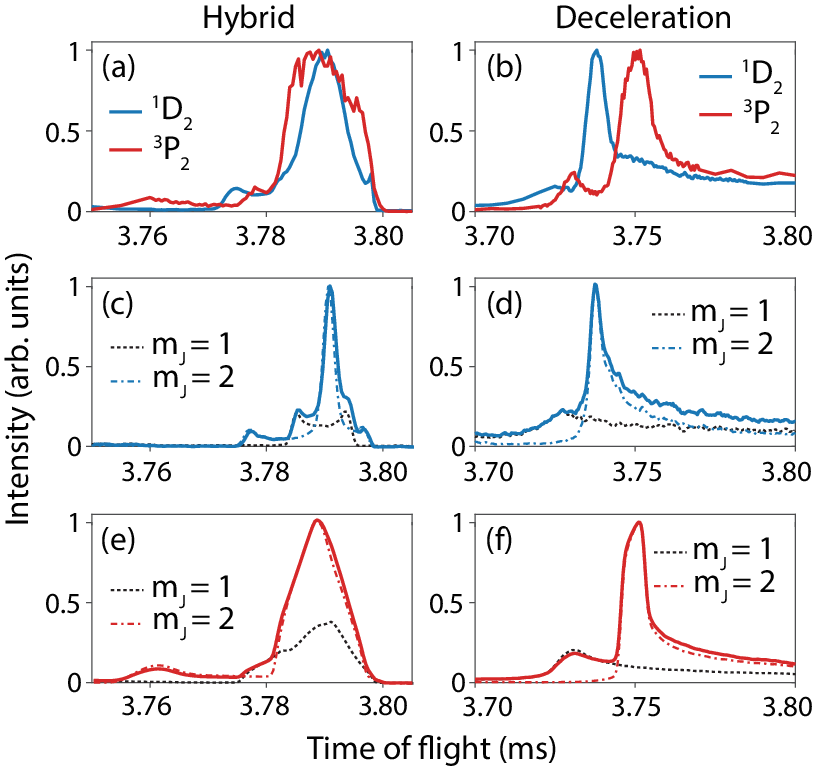}
	\caption{Selected parts of measured TOF profiles for \SoneDJ (blue) and \SthreePJ{2} (red) atoms exiting the Zeeman decelerator, operated in hybrid (a) or deceleration mode (b), both with a mean forward velocity of 850~m/s. Each profile is normalized to its peak intensity to facilitate comparison. Simulated TOF profiles for \SoneDJ are shown in (c) and (d), and for \SthreePJ{2} in (e) and (f), corresponding to the experimental conditions in (a) and (b), respectively. The contributions of the low-field-seeking $m_J = 1$ and $m_J = 2$ states to the simulated profiles are indicated by dashed and dash-dotted lines, respectively.}
	\label{fig:TOFs_850decvshyb_allmeasvssim}
\end{figure}

To demonstrate this, we operated the decelerator in two modes to reach the same final velocity of 850 m/s: hybrid mode to guide at a constant velocity, and deceleration mode starting from an initial velocity of 875 m/s. Although both pulse sequences are optimized for selecting \SoneDJ atoms in the $m_J = 2$ state, other low-field-seeking states of both the $^1D_2$ and $^3P_J$ states are transmitted through the decelerator as well. Figs.~\ref{fig:TOFs_850decvshyb_allmeasvssim}a and b show selected parts of the corresponding TOF profiles recorded for \SoneDJ (blue) and \SthreePJ{2} (red) atoms for the hybrid and deceleration mode, respectively. Each profile is normalized to its peak intensity to facilitate comparison. 

In hybrid mode, the \SoneDJ and \SthreePJ{2} packets exit the decelerator simultaneously, and it is not possible to separate the atoms in the two quantum states. However, in deceleration mode, the \SthreePJ{2} packet arrives later than the \SoneDJ packet due to its larger magnetic moment and therefore stronger interaction with the magnetic fields. By selecting the arrival time corresponding to the peak of the \SoneDJ packet, the quantum-state purity of the packet can be significantly improved. The purity could be further enhanced by employing a heavier seed gas, which lowers the beam velocity and increases the effectiveness of deceleration, thereby improving the temporal separation between the different states. This approach, however, would also reduce the overall \SoneDJ density, as discussed in section \ref{sec:TOF}. Note that, in a similar way, a high-purity \SthreePJ{2} packet could be produced as well.

The measured TOF profiles show good agreement with the corresponding simulated profiles, shown in Figs.~\ref{fig:TOFs_850decvshyb_allmeasvssim}c and d for the \SoneDJ state and in Figs.~\ref{fig:TOFs_850decvshyb_allmeasvssim}e and f for \SthreePJ{2}, although some experimental features are slightly broader or shifted. In all profiles, multiple peaks are observed, originating from different low-field-seeking $m_J$ states with distinct magnetic moments. This is illustrated by the simulated $m_J$-state-resolved TOF profiles, shown by the dashed and dash-dotted lines in Fig.~\ref{fig:TOFs_850decvshyb_allmeasvssim} for $m_J = 1$ and $m_J = 2$, respectively . The $m_J = 0$ state is not shown, since it has no interaction with the magnetic field and therefore is negligible in intensity. High-field-seeking $m_J = -1, -2$ states are defocused by the hexapoles in the decelerator, and therefore do not reach the detection region. 

The $m_J = 2$ level of the $^3P_2$ state has the highest magnetic moment ($3~\mu_B$). It is therefore decelerated the most and arrives latest at a mean TOF of 3.751 ms (Fig.~\ref{fig:TOFs_850decvshyb_allmeasvssim}f). The $m_J = 2$ level of the $^1D_2$ state, with a magnetic moment of $2~\mu_B$, arrives slightly earlier at a mean TOF of 3.737 ms (Fig.~\ref{fig:TOFs_850decvshyb_allmeasvssim}d). The $m_J = 1$ levels of the $^1D_2$ and $^3P_2$ states, with magnetic moments of $1.5~\mu_B$ and $1~\mu_B$, respectively, result in lower-intensity peaks at earlier arrival times due to their lower interaction with the magnetic field. This demonstrates the decelerator's ability to temporally resolve quantum states of S atoms based on their magnetic moments. It should be noted, however, that in our experimental setup, the $m_J$ sublevels of a given $J$ state become degenerate and redistribute before detection due to an insufficient magnetic field between the end of the decelerator and the detection region. As a result, they cannot be distinguished experimentally.

\subsection{Elastic scattering}
Estimating particle densities from the observed signal levels in our experiment is particularly challenging, and we therefore refrain from this. Instead, to assess whether the beam intensity is sufficient for scattering studies, we performed a proof-of-principle experiment involving elastic collisions between \SoneDJ and Ar atoms. 

\begin{figure}[H]
	\centering
	\includegraphics[width=\columnwidth]{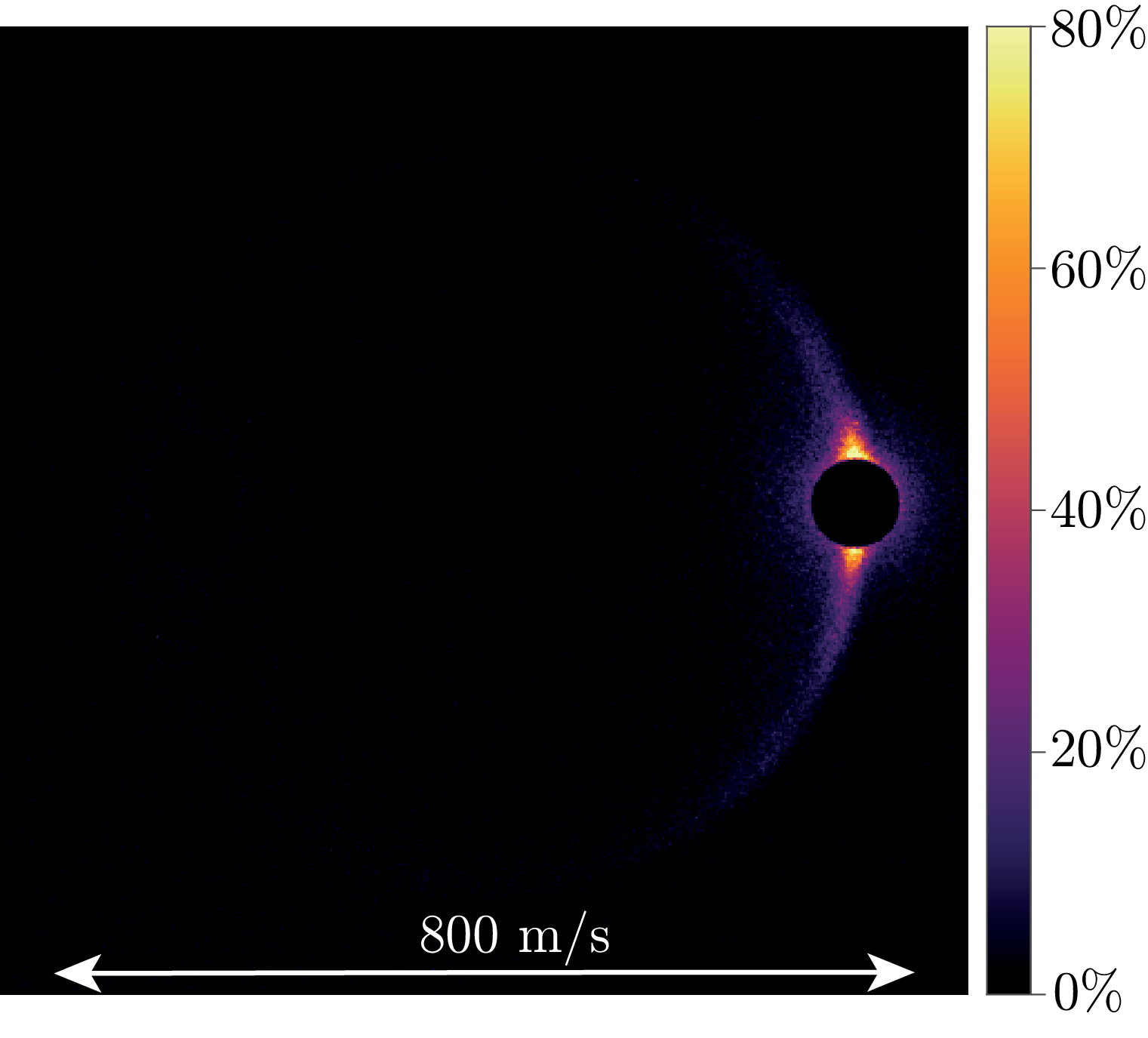}
	\caption{Experimental elastic scattering image for S($^1D_2$) + Ar. Part of the image is masked where the initial \SoneDJ beam contributes to the signal. }
	\label{fig:TElastic}
\end{figure}

In this experiment, the \SoneDJ atoms were guided through the decelerator in hybrid mode to a final velocity of 875 m/s and intersected a pure Ar beam with a velocity of around 610 m/s at an intersection angle of 45$^\circ$. This beam was produced by expanding 3 bar Ar through an Even-Lavie valve \cite{Even:AiC2014:636042}, operated at a repetition rate of 20 Hz. This resulted in a collision energy of about 285~\invcm. Fig. \ref{fig:TElastic} shows the resulting VMI image of the scattered \SoneDJ atoms, recorded using the 1+1 REMPI detection scheme described in the Experimental section. In the image, the relative velocity vector is oriented horizontally, with forward scattering appearing on the right-hand side. The observed signal levels were sufficient to clearly resolve the scattering distribution with a favorable signal-to-noise ratio. On average, we detected around 1-2 events per laser shot by reducing the laser power to only 55 $\mu$J, demonstrating that the \SoneDJ beam produced by our decelerator is intense enough for scattering studies. 

\section{Conclusions}
We have demonstrated the successful production of a high-intensity and velocity-controlled beam of electronically excited \SoneDJ atoms using our multistage Zeeman decelerator. Following photolysis of CS$_2$, both \SoneDJ and ground-state \SthreePJ{J} atoms are formed, and both can be manipulated using the decelerator. By operating the decelerator in deceleration mode, we achieved temporal separation between \SoneDJ and \SthreePJ{2} atoms, enabling the generation of a S-atom beam with enhanced quantum-state purity when required. 

Furthermore, we showed that the density of the \SoneDJ beam is sufficient to perform elastic scattering experiments, as evidenced by our VMI image obtained for \SoneDJ + Ar collisions. This is an excellent starting point for investigating reactive scattering 
between \SoneDJ and H$_2$ (including its isotopologues), as well as electronic quenching of \SoneDJ in collisions with various partners. Since cross sections for such processes could be one to two orders of magnitude smaller than those for elastic scattering, their measurement will be considerably more challenging. Nevertheless, the precise control over collision energy provided by our setup would make it well suited to explore these processes under well-defined conditions.

\section{Acknowledgements}
We are grateful to S.Y.T. van de Meerakker for his continued support and for many valuable discussions. We thank X.-D. Wang, S. Kusters, and B. Vonk for their contributions to preparatory studies, and A. van Roij, E. Sweers, and G. Wulterkens for their expert technical support. 
This work is part of the research program of the Dutch Research Council (NWO). J.O. acknowledges support from the European Research Council (IQ-SCORES, Grant Agreement No. 101160597). 

\section{Data availability}
The data that support the findings of this article are openly available \cite{Tsoukala:data:2025}.

\end{document}